\documentclass[3p,12pt]{elsarticle}

\usepackage[numbers]{natbib}
\usepackage{amsmath,amssymb,amsfonts}
\usepackage{graphicx}
\usepackage{booktabs}
\usepackage{url}
\usepackage{hyperref}
\usepackage{placeins}

\journal{Speech Communication}

\begin{document}

\begin{frontmatter}

\title{Half-Truth Audio Detection and Localisation: A Lightweight
Cross-Attentive Architecture and a Cross-Corpus Diagnostic Study}

\author[inst1]{S Sutharya\corref{cor1}}
\ead{sutharya8@gmail.com}
\author[inst1]{Remya K Sasi}
\cortext[cor1]{Corresponding author}
\address[inst1]{Department of Computer Science, Cochin University of Science and Technology, India}

\begin{abstract}
Partially manipulated (half-truth) speech, where a short synthesised
segment is spliced into an otherwise genuine utterance, is a harder and
more realistic forensic threat than the fully-synthesised deepfakes that
dominate the literature. We present CAFNet, a lightweight (576K-parameter,
2.24~MB) cross-attentive architecture that fuses MFCC, LFCC, and
Chroma-STFT features to jointly classify audio as real, fully fake, or
half-truth, and regress the temporal boundaries of the synthesised
region, at approximately 14~ms CPU latency. A component ablation shows
cross-attention fusion is CAFNet's most load-bearing component; a deeply
supervised auxiliary classification head from earlier iterations is not,
and removing it improves every in-domain metric under 3-seed replication
with substantially lower variance. On MLADDC T2+T3 the model reaches 97.55\%$\pm$0.69\% ternary accuracy and 0.037~s boundary mean absolute error (MAE), to our knowledge, the first reported continuous splice- boundary localisation result on this benchmark. Zero-shot evaluation on two  independent benchmarks shows transfer is capability- and corpus-dependent rather than uniform: on Half-Truth Audio Detection dataset (HAD), detection recall reaches 84.9\% and ternary classification resolves half-truth correctly on half of true half-truth clips (50.4\%), while on PartialSpoof, binary detection stays near chance (AUC 0.5544). We treat this asymmetry, not a single generalization verdict, as the finding. HAD localisation improves in absolute terms but degrades in relative terms, since in-domain localisation improved faster. An architectural change validated purely in-domain thus shifted the cross-corpus transfer profile, evidence that cross-corpus evaluation should accompany, not follow, in-domain architecture decisions.
\end{abstract}


\end{frontmatter}

\section{Introduction}
\label{sec:intro}

Audio deepfake detection has matured quickly around a single framing: given
an utterance, decide whether it is real or synthesised. That framing does
not match how manipulated audio is actually used to deceive. A forged
voicemail, a doctored political statement, or a manipulated
customer-service call rarely replaces an entire recording; it replaces a
few seconds of it, spliced into speech that is otherwise genuine. Half-truth
audio of this kind is disproportionately dangerous relative to its size:
75\% of the signal is unmodified real speech, so listeners and detectors
alike default to trusting the whole clip, while the altered fragment can
invert the meaning of what was said. Controlled listening studies confirm
humans cannot reliably catch such splices \citep{mai2023}, and neural
vocoders such as HiFi-GAN \citep{hifigan2020} and BigVGAN
\citep{bigvgan2023} now produce segments spectrally indistinguishable from
natural speech even under close listening. Detector robustness under
out-of-distribution and adversarial conditions, rather than in-domain
accuracy alone, is an increasingly active concern
\citep{rabhi2024,dixit2023,kawa2023}; we examine natural domain shift, not
adversarial perturbation, and treat the two as distinct threat models.

Detection and localisation are distinct capabilities. Detection asks
whether a clip contains a manipulation at all; localisation asks where.
A system that only answers the first question is of limited forensic
use, since flagging a clip as suspicious does not tell an investigator,
moderator, or downstream verification pipeline which few seconds of a
long recording to examine. The literature is nonetheless organised almost
entirely around the first question; prior work on partially spoofed audio
explores localisation alongside classification
\citep{halftruth2021,cai2024boundary,zhang2022partialspoof}, but almost
always as a secondary or frame-level segmentation task on a single
benchmark. Joint ternary classification (real / fully-fake / half-truth)
with continuous-boundary regression, evaluated across benchmarks, remains
addressed by comparatively few studies.

This paper has three contributions. First, a joint architecture: CAFNet combines 
multi-feature handcrafted fusion (in the spirit of MFAAN \citep{mfaan2023}) with
cross-attention fusion and a BiLSTM boundary-regression head to produce
coupled outputs, a coarse detection signal, a three-way class, and a
continuous boundary estimate, from a single lightweight forward pass.
Second, a component ablation of that architecture's own claims, including
a deeply supervised auxiliary classification head whose benefit was never
isolated in the original design. Under a training protocol that
checkpoints on best validation loss with early stopping rather than a
fixed epoch budget, the auxiliary head turns out not to be load-bearing:
removing it improves in-domain accuracy, EER, and localisation MAE
simultaneously (Section~\ref{sec:cafablation}). Third, and the paper's
main contribution: because CAFNet's outputs are produced jointly, they
can be evaluated separately under domain shift. Applying the corrected,
auxiliary-head-free checkpoint zero-shot to two corpora it never saw
during training, HAD \citep{halftruth2021} and PartialSpoof
\citep{zhang2022partialspoof}, and comparing against an identical
evaluation of the original architecture, shows the two are not equivalent
under domain shift even though the architectural change was validated
purely on in-domain metrics: detection and ternary classification
transfer substantially better under the corrected architecture, while
the localisation head's relative out-of-domain degradation widens rather
than narrows, because it improved faster in-domain than out-of-domain
(Section~\ref{sec:generalisation}).

Section~\ref{sec:related} reviews related work. Section~\ref{sec:method}
describes the dataset, features, and architecture. Section~\ref{sec:results}
reports in-domain results, including the component ablation.
Section~\ref{sec:generalisation} presents the cross-dataset study.
Section~\ref{sec:conclusion} states limitations and concludes.

\section{Related Work}
\label{sec:related}

Binary deepfake detection began from handcrafted spectral features.
Khochare et al.\ \citep{khochare2021} reported 67\% accuracy on the FoR
dataset using MFCCs with an SVM, and Hamza et al.\ \citep{hamza2022} paired
MFCCs with a VGG16 backbone on the same corpus. LFCC, which uses a uniform
rather than Mel-warped filterbank, is more sensitive to the high-frequency
artefacts neural vocoders leave behind; the MLADDC baseline reports LFCC
with a CNN as its strongest single-feature system (68.44\% accuracy,
40.9\% EER on the binary T2 track) \citep{mladdc2024}. MFAAN
\citep{mfaan2023} combines MFCC, LFCC, and Chroma-STFT under the premise
that synthesis artefacts show up differently across timbral, spectral, and
harmonic axes; it is a binary classifier by construction, with no path to
a third class or a regression output. We adopt the same three-feature
philosophy but treat it as a starting point for the fusion mechanism
itself (Section~\ref{sec:results}). Self-supervised front-ends such as XLS-R 
\citep{xlsr2022} and the Audio Spectrogram Transformer \citep{ast2021}, built on 
wav2vec-2.0-style representations \citep{baevski2020}, have driven the strongest 
reported binary results on ASVspoof \citep{asvspoof2021}. These models report
state-of-the-art accuracy at the cost of tens to hundreds of millions of
parameters and correspondingly heavy inference budgets, which sits at
odds with deployment on the resource-constrained endpoints (mobile
devices, embedded moderation pipelines, on-device call screening) where
splice localisation is often most needed. CAFNet targets this
complementary design point: a three-way, boundary-regression-capable
model under 600K parameters and low-double-digit-millisecond CPU
latency, rather than pooled binary accuracy at SSL-backbone scale.

Binary framing does not capture how manipulated audio is actually
deployed. Yi et al.\ introduced Half-Truth (HAD) \citep{halftruth2021}
specifically to study partially fake audio, and established it is
substantially harder to detect than fully fake audio. Zhang et al.'s
PartialSpoof \citep{zhang2022partialspoof,zhang2021initial} independently
built a partially-spoofed benchmark on ASVspoof 2019
\citep{wang2020asvspoof2019} with fine-grained, multi-resolution
segment-level labels and explicit voice-activity annotation, a design
choice we return to in Sections~\ref{sec:dataset-justification}
and~\ref{sec:detect}. MLADDC \citep{mladdc2024} extends the half-truth
setting to a multilingual scale (20 languages) with a fixed single-splice
protocol. Across all three corpora, published baselines report
classification metrics almost exclusively; boundary localisation is
addressed by comparatively little work. Cai and Li \citep{cai2024boundary}
explore frame-level boundary detection but do not evaluate on MLADDC T3
or study localisation jointly with multilingual ternary classification.
This is the gap CAFNet targets.

Cross-corpus generalisation in deepfake detection is known to be poor:
M\"{u}ller et al.\ \citep{mueller2022} showed binary detectors trained on
one corpus degrade sharply on another, a result we confirm again in
Table~\ref{tab:zeroshot}. Recent work targets SSL-based cross-dataset
robustness specifically: Zhang et al.\ \citep{zhang2024sls} pair XLS-R
with a dedicated classifier head for improved cross-dataset performance;
El Kheir et al.\ \citep{elkheir2025} analyse which SSL layers carry
spoofing-relevant information across corpora; Jung et al.'s SpoofCeleb
\citep{jung2025spoofceleb} targets in-the-wild evaluation. These studies
ask which features transfer; we ask which of a jointly-trained model's
outputs (detection, classification, localisation) transfer, holding the
representation fixed.

\section{Methodology}
\label{sec:method}

\subsection{Dataset and Justification}
\label{sec:dataset-justification}

We train and evaluate in-domain on MLADDC \citep{mladdc2024}, which
contains 400,000 recordings across 20 languages: 80,000 real utterances,
160,000 fully synthesised deepfakes (T2), and 160,000 half-truth clips
(T3). We chose MLADDC over HAD and PartialSpoof for scale and linguistic
diversity, for its single continuous splice interval per half-truth clip
(mapping directly onto boundary regression, versus HAD/PartialSpoof's
segment- or frame-level labels), and for an explicit ternary protocol at
the track level.

CAFNet's input is a fixed 4~s window; MLADDC's source recordings are not
uniformly 4~s, so a window is selected from each recording at both
training and test time. For T2 this is a single fixed 4~s crop per file.
For T3, the $\sim$1~s synthesised segment must fall inside whichever
window is used, or the boundary-regression target is undefined; the
window's start offset is drawn uniformly at random from every position
for which the full splice fits inside a 4~s window, identically at
training and in-domain test time. This defines the corpus's native input
distribution rather than granting the test set information the training
set lacked, and differs from the oracle windowing used for zero-shot
localisation on HAD (Section~\ref{sec:oodloc}), where the same placement
rule is applied at test time only, using HAD's ground-truth timestamps,
to a checkpoint that never learned HAD's own splice-position
distribution.

MLADDC provides no explicit non-speech annotation, unlike PartialSpoof,
which ships three-state voice-activity labels; CAFNet has no mechanism to
separate speech from non-speech content inside its window. Silence and
non-speech intervals are a documented source of spurious discriminative
signal in ASVspoof-derived corpora \citep{muller2021silence,zhang2023silence}.

The combined T2+T3 training split contains 262,400 samples (44,800 real /
89,600 fully fake / 128,000 half-truth); validation and test splits each
contain 32,800 samples in the same proportions.

\subsection{Feature Extraction}

Audio is resampled to 16~kHz mono. Three feature representations are
extracted in parallel (Table~\ref{tab:features}). A correct LFCC
implementation requires an explicit linear-spaced filterbank; applying
librosa's \texttt{mfcc} function to a spectrogram introduces Mel warping
and does not yield true LFCC, so the filterbank is implemented separately.
During training, standard augmentation is applied stochastically:
temporal masking (10--30 frames, $p{=}0.3$), frequency masking (2--8
rows, $p{=}0.3$), and additive Gaussian noise ($\sigma{=}0.01$,
$p{=}0.15$).

\begin{table}[h]
  \centering
  \caption{Feature representations extracted from each audio clip.}
  \label{tab:features}
  \begin{tabular}{llll}
    \toprule
    \textbf{Feature} & \textbf{Coeff.} & \textbf{Shape} & \textbf{Captures} \\
    \midrule
    MFCC        & 40 & $40{\times}251$ & Timbral texture (Mel) \\
    LFCC        & 40 & $40{\times}251$ & High-freq.\ artefacts \\
    Chroma-STFT & 12 & $12{\times}251$ & Harmonic pitch-class  \\
    \bottomrule
  \end{tabular}
\end{table}

\subsection{MFAAN Baseline}
\label{sec:mfaan}

MFAAN \citep{mfaan2023} is re-implemented as a 2D-CNN binary baseline
following the original paper: two Conv2d--ReLU--Dropout--MaxPool2d blocks
(128 channels) per feature path, adaptive average pooling, concatenation,
and a two-layer dense head with a two-class output (322,562 parameters).
MFAAN has no third-class output and no regression path, so it cannot
address T3 ternary classification or localisation; it serves here as a
feature-fusion reference point for Section~\ref{sec:ablation}.

\subsection{CAFNet Architecture}
\label{sec:cafnet}

\begin{figure}[h]
  \centering
  \includegraphics[width=\columnwidth]{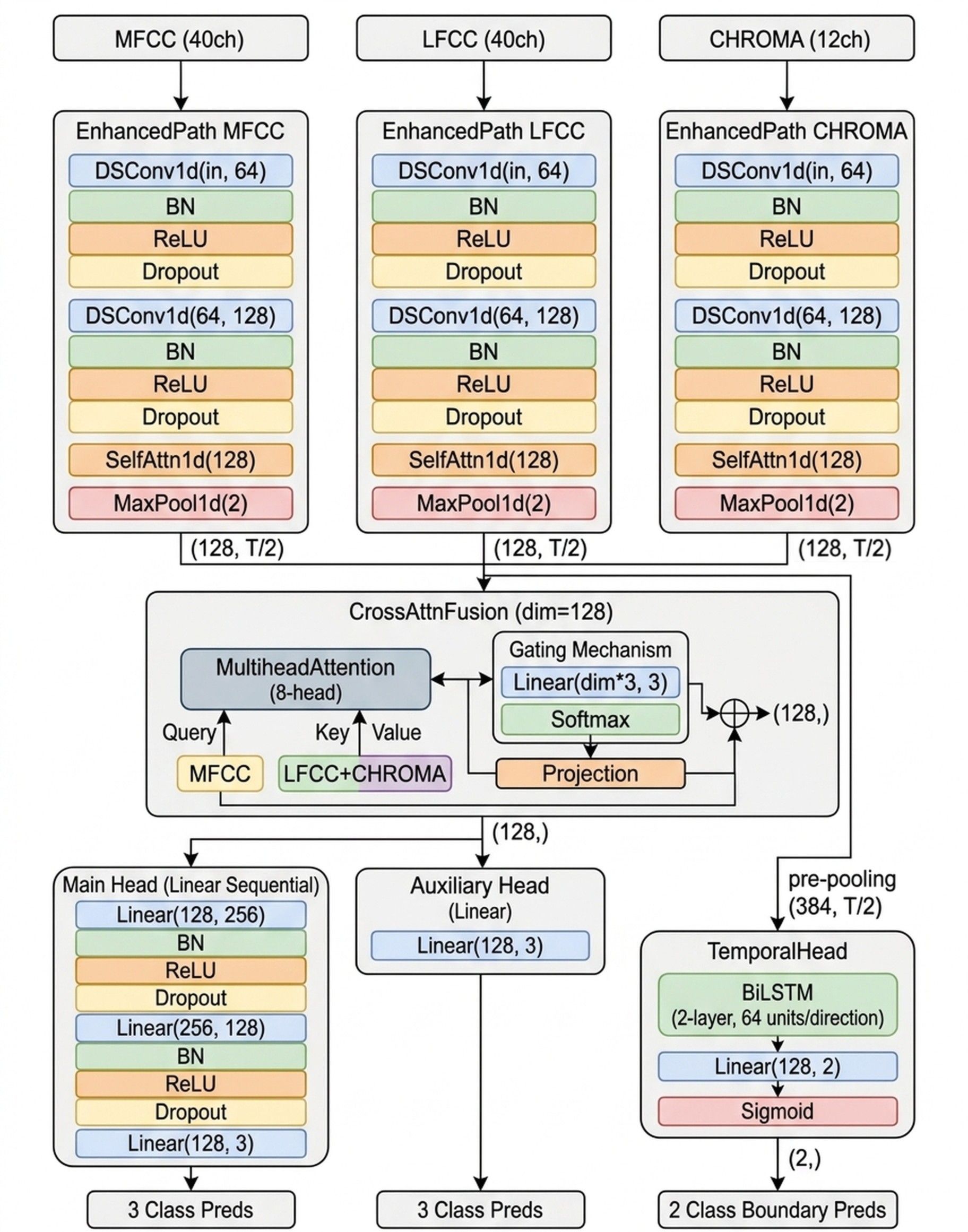}
  \caption{CAFNet architecture (canonical, auxiliary-head-free form): Three
  EnhancedPath encoders feed a cross-attention fusion module; the fused
  representation drives a single three-class classification head, and the
  concatenated pre-pooling path outputs drive a BiLSTM
  temporal-boundary regression head.}
  \label{fig:cafnet}
\end{figure}

CAFNet retains MFAAN's parallel multi-feature design but replaces 2D-CNN
pooling with 1D temporal convolutions, adds cross-attention fusion,
extends classification to three classes, and adds a BiLSTM regression
path for splice-boundary prediction (Figure~\ref{fig:cafnet}). The description 
below is CAFNet's canonical form, arrived at via the component ablation in 
Section~\ref{sec:cafablation}, which identifies cross-attention fusion as load-
bearing and a deeply supervised auxiliary classification head as unnecessary.

\textbf{EnhancedPath.} Each feature matrix is treated as a 1D temporal
sequence through two depthwise-separable convolution blocks (depthwise
Conv1d, pointwise Conv1d, BatchNorm, ReLU, Dropout), expanding channels
from the input dimension to 64 then 128. A lightweight self-attention
module (query/key projections of dimension 16) refines the sequence,
added residually through a learnable scalar initialised to zero.
MaxPool1d(2) halves the temporal dimension.

\textbf{CrossAttentionFusion.} The three EnhancedPath outputs are fused via
cross-attention: the MFCC sequence serves as query, the concatenated LFCC
and Chroma sequences as key-value input to an 8-head MultiheadAttention
layer (dimension 128). The three mean-pooled path outputs are additionally
combined through a learned gating mechanism before a final linear
projection to 128-d. Section~\ref{sec:ablation} shows naive concatenation
of Chroma-STFT under MFAAN's simpler fusion \emph{hurts} accuracy;
cross-attention avoids this, and Section~\ref{sec:cafablation} shows it is
CAFNet's single most load-bearing component.

\textbf{Classification and Temporal Heads.} A two-layer dense head produces
three-class logits. The pre-pooling EnhancedPath outputs are concatenated
and passed to a two-layer bidirectional LSTM (64 units per direction)
followed by a sigmoid-activated linear layer predicting normalised start
and end boundaries in $[0,1]$. Total parameters: \textbf{576,027}.

\FloatBarrier

\subsection{Training}
\label{sec:training}

MFAAN is trained on MLADDC T2 with batch size 64, AdamW
($\mathrm{lr}{=}5{\times}10^{-4}$, weight decay $10^{-4}$), weighted
cross-entropy (class weights 2.0/1.0), and ReduceLROnPlateau with early
stopping (patience 10).

CAFNet uses batch size 64, AdamW ($\mathrm{lr}{=}5{\times}10^{-4}$),
gradient clipping to norm 1.0, and a composite loss:
\begin{equation}
  \mathcal{L} = \mathcal{L}_{\mathrm{cls}} + 0.3\,\mathcal{L}_{\mathrm{temp}},
  \label{eq:loss}
\end{equation}
where $\mathcal{L}_{\mathrm{cls}}$ is three-class weighted cross-entropy
(weights 1.622 / 0.811 / 0.568 for real / fake / half-truth) and
$\mathcal{L}_{\mathrm{temp}}$ is MSE over normalised boundaries, computed
only for half-truth samples. The composite loss omits a term for a deeply 
supervised auxiliary head; Section~\ref{sec:cafablation}'s component ablation 
shows this component does not improve performance under a best-validation-loss 
training protocol.

Headline results use a protocol that checkpoints on best validation loss
with early stopping (patience 7, capped at 20--25 epochs);
Section~\ref{sec:cafablation} shows a fixed, low epoch-budget alternative
under-trains variants unevenly and can reverse which variant looks best.
The canonical checkpoint carried through the paper is the
auxiliary-head-free architecture at seed 42, the lowest-validation-loss
run (0.0603) among three seeds (Section~\ref{sec:cafablation}); every
localisation and cross-dataset result (Section~\ref{sec:generalisation})
uses this single checkpoint, applied zero-shot with no fine-tuning on HAD
or PartialSpoof.

\subsection{Evaluation Protocol}
\label{sec:evalprotocol}

Accuracy, EER (Brent's method on the ROC curve), and AUC (macro
one-versus-rest for three-class; standard binary otherwise) are reported.
For three-class CAFNet in binary contexts, EER is computed as real versus
non-real. Temporal MAE is converted to seconds by multiplying normalised
error by 4.0. For diagnostic purposes only, we additionally swept the decision
threshold over a 99-point grid on the PartialSpoof evaluation set itself
(Section~\ref{sec:detect}); the resulting best-case accuracy is a
threshold-oracle upper bound, not a deployable operating point, and is
not reported as a headline metric. AUC, EER,
and the paired TPR/FPR reported alongside it are threshold-independent or
fixed-threshold ($0.5$) and carry the evidentiary weight there.
In-domain windowing follows Section~\ref{sec:dataset-justification},
identically at train and test time. Out-of-domain detection and
classification (HAD, PartialSpoof) crop or pad each utterance to 4~s from
its beginning, with no privileged splice-location information.
Out-of-domain \emph{localisation} on HAD (Section~\ref{sec:oodloc}) uses
ground-truth-informed windows, an oracle protocol justified there.

\section{In-Domain Results on MLADDC}
\label{sec:results}

All experiments are conducted on Kaggle's compute environment using dual
NVIDIA T4 GPUs with PyTorch 2.x.

\begin{table}[h]
  \centering
  \caption{Model architecture and training configuration (canonical,
  auxiliary-head-free CAFNet).}
  \label{tab:modelparams}
  \begin{tabular}{lll}
    \toprule
    \textbf{Component} & \textbf{CAFNet} & \textbf{MFAAN} \\
    \midrule
    Parameters      & 576,027              & 322,562 \\
    Encoder         & DSConv1d             & Conv2d  \\
    Channels        & $64{\to}128$         & 128     \\
    Fusion          & Cross-attn (8-head)  & Concat  \\
    Classif.\ head  & 3-class (single head) & 2-class \\
    Temporal head   & BiLSTM ($64{\times}2$) & None  \\
    Checkpoint size & 2.24~MB & --- \\
    CPU latency (4~s clip, mean) & 14.4~ms & --- \\
    \bottomrule
  \end{tabular}
\end{table}

\subsection{Component Ablation: Justifying the Auxiliary Head Removal}
\label{sec:cafablation}

The original architecture combined three components never isolated from
one another: cross-attention fusion, self-attention refinement, and a
deeply supervised auxiliary classification head. We ablate each on the
unified T2+T3 task, CAFNet's actual training objective, training every
variant to its best validation-loss checkpoint with early stopping
(patience 7, cap 25 epochs) rather than a fixed epoch budget, since
short fixed-epoch training leaves slower-converging variants
(cross-attention fusion has more parameters to learn than simple
concatenation) mid-convergence and produces an unstable ranking. Two
variants (no self-attention; all three components removed) trail the
remaining three by a wide, consistent margin under this protocol and are
omitted from Table~\ref{tab:ablation-round2} for brevity, which reports
the three variants whose ranking is not obvious a priori.

\begin{table}[h]
  \centering
  \caption{CAFNet component ablation, best-validation-loss checkpoint with
  early stopping (unified T2+T3 test set, $n{=}32{,}800$, seed 42).}
  \label{tab:ablation-round2}
  \begin{tabular}{lllll}
    \toprule
    \textbf{Variant} & \textbf{Acc} & \textbf{AUC} & \textbf{EER} & \textbf{tMAE} \\
    \midrule
    Full CAFNet         & 97.61\% & 0.9995 & 1.21\% & 0.029s \\
    No cross-attention  & 96.02\% & 0.9991 & 1.57\% & 0.084s \\
    No auxiliary head   & \textbf{98.04\%} & \textbf{0.9995} & \textbf{1.23\%} & 0.035s \\
    \bottomrule
  \end{tabular}
\end{table}

Cross-attention fusion is clearly load-bearing: removing it costs 1.6
accuracy points and nearly triples boundary-localisation error
(0.029~s $\to$ 0.084~s), showing the fusion mechanism matters for the
shared representation the temporal head reads from, not just
classification. The auxiliary head is not load-bearing: removing it
improves accuracy, EER, and AUC simultaneously, with tMAE (0.035~s)
within ordinary run-to-run noise of the full architecture's 0.029~s.

To verify this is not a single-run artefact, we replicate the full and
auxiliary-head-free variants over three seeds (42, 123, 2024; 20-epoch
cap, same protocol). Table~\ref{tab:multiseed} confirms the direction on
every metric's mean, with roughly one third the run-to-run standard
deviation of the full architecture. We adopt the auxiliary-head-free
architecture as canonical; the checkpoint carried forward into every
remaining result is its seed-42 run, the lowest-validation-loss
checkpoint (0.0603) among all six runs above.

\begin{table}[h]
  \centering
  \caption{Multi-seed replication, full vs.\ auxiliary-head-free CAFNet
  ($n{=}3$ seeds: 42, 123, 2024; unified T2+T3 test set).}
  \label{tab:multiseed}
  \begin{tabular}{llll}
    \toprule
    \textbf{Variant} & \textbf{Acc (mean$\pm$std)} & \textbf{EER (mean$\pm$std)} & \textbf{tMAE (mean$\pm$std)} \\
    \midrule
    Full             & 95.80\%$\pm$1.84\% & 3.46\%$\pm$2.14\% & 0.045s$\pm$0.023s \\
    No auxiliary head & \textbf{97.55\%$\pm$0.69\%} & \textbf{1.37\%$\pm$0.43\%} & \textbf{0.029s$\pm$0.007s} \\
    \bottomrule
  \end{tabular}
\end{table}

\subsection{Half-Truth Localisation (In-Domain)}
\label{sec:temporal}

Table~\ref{tab:temploc} reports boundary localisation for the canonical
checkpoint across all 16,000 MLADDC half-truth test clips. Median overall
error is 0.027~s; 99.1\% of predictions fall within 0.25~s. To our
knowledge, no prior published work reports a continuous boundary
localisation result on MLADDC's half-truth track.

\begin{table}[h]
  \centering
  \caption{Temporal boundary localisation on MLADDC T3 (16,000 clips,
  canonical checkpoint).}
  \label{tab:temploc}
  \begin{tabular}{lllll}
    \toprule
    \textbf{Boundary} & \textbf{MAE (s)} & \textbf{Median (s)} & \textbf{p90 (s)} & \textbf{$\le$0.25s} \\
    \midrule
    Start   & 0.040 & 0.028 & 0.075 & 0.992 \\
    End     & 0.034 & 0.023 & 0.061 & 0.991 \\
    Overall & 0.037 & 0.027 & 0.060 & 0.991 \\
    \bottomrule
  \end{tabular}
\end{table}

\begin{figure}[h]
  \centering
  \includegraphics[width=\columnwidth]{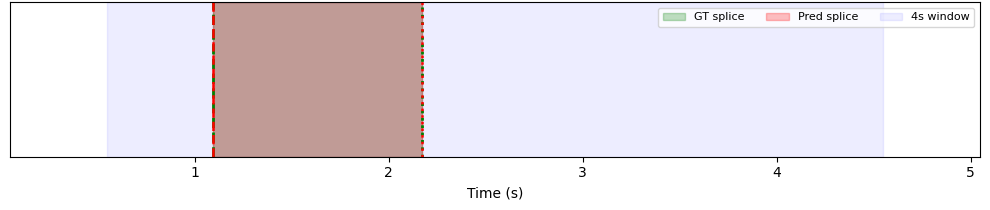}
  \caption{Best case (\texttt{swedish\_3875\_hifi}), canonical checkpoint:
  predicted boundaries (dashed) coincide exactly with the ground-truth
  splice (shaded).}
  \label{fig:locbest}
\end{figure}

\begin{figure}[h]
  \centering
  \includegraphics[width=\columnwidth]{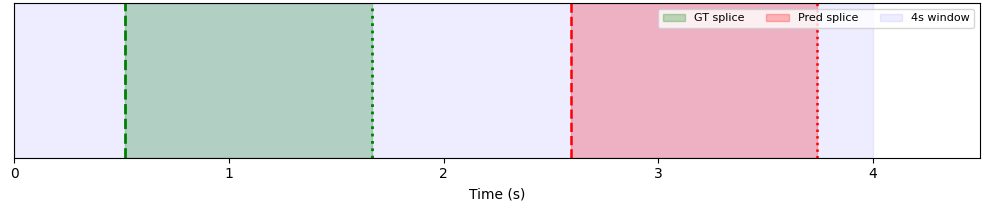}
  \caption{Hard case (\texttt{spanish\_330\_hifi}), canonical checkpoint:
  the predicted interval (dashed) diverges sharply from the ground truth
  (shaded); the clip is also misclassified as real.}
  \label{fig:locworst}
\end{figure}

Localisation error correlates with classification correctness on the same
clips: 15,327 of 16,000 (95.8\%) are correctly classified half-truth, with
MAE 0.034~s on that subset; the 673 misclassified clips (4.2\%) have MAE
0.105~s. Figure~\ref{fig:locbest} shows a representative best case
(\texttt{swedish\_3875\_hifi}: ground truth 1.094--2.171~s, predicted
1.094--2.171~s, zero error, $p(\mathrm{HT}){=}1.000$).
Figure~\ref{fig:locworst} shows a hard case
(\texttt{spanish\_330\_hifi}: ground truth 0.515--1.666~s, predicted
2.592--3.742~s, combined error 2.076~s), where the clip is also
misclassified as real ($p(\mathrm{HT}){=}0.201$): the localisation and
classification failures co-occur, a pattern that recurs, far more
sharply, out-of-domain (Section~\ref{sec:ternary}).

\subsection{Three-Class Detection}
\label{sec:unified}

Table~\ref{tab:unified} reports the multi-seed mean$\pm$std as the
headline accuracy/EER/tMAE; AUC is the single seed-42 figure, since AUC
was not aggregated across the replication. The original MLADDC authors
report at most 58.01\% testing accuracy on T3 (LFCC+BiLSTM, binary
framing) \citep{mladdc2024}; CAFNet's ternary accuracy on the combined
track is substantially higher, though not directly comparable given the
different framing.

\begin{table}[h]
  \centering
  \caption{CAFNet unified three-class performance on MLADDC T2+T3
  (canonical, auxiliary-head-free architecture).}
  \label{tab:unified}
  \begin{tabular}{ll}
    \toprule
    \textbf{Metric} & \textbf{Value} \\
    \midrule
    Overall accuracy (3-seed mean$\pm$std) & 97.55\%$\pm$0.69\% \\
    EER, real vs.\ non-real (3-seed mean$\pm$std) & 1.37\%$\pm$0.43\% \\
    Temporal MAE, overall (3-seed mean$\pm$std) & 0.029s$\pm$0.007s \\
    Macro AUC, OvR (seed 42 only)  & 0.9995  \\
    \bottomrule
  \end{tabular}
\end{table}

\begin{figure}[h]
  \centering
  \includegraphics[width=\columnwidth]{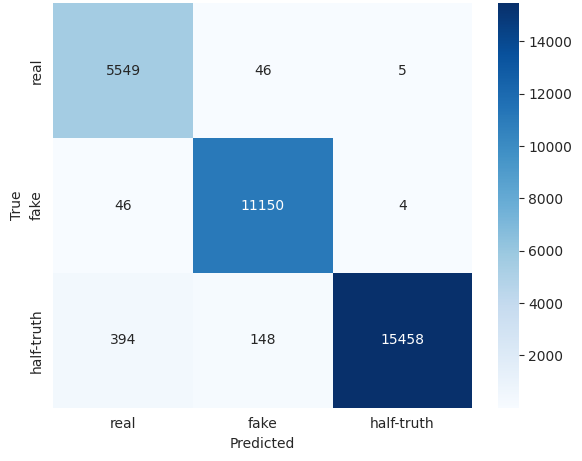}
  \caption{Confusion matrix for CAFNet three-class prediction on the
  MLADDC T2+T3 test set (32,800 clips), canonical checkpoint.}
  \label{fig:confmat}
\end{figure}

Table~\ref{tab:perclass} gives per-class detail for the seed-42 checkpoint
(accuracy 98.04\%, matching Table~\ref{tab:ablation-round2}), with the
confusion matrix in Figure~\ref{fig:confmat}. Half-truth achieves
near-perfect precision (0.9994) at a modest recall cost (0.9661): the
model rarely raises a false half-truth alarm but misses some true
half-truth clips as something else. Real remains the weakest class, but
its precision improves sharply over the earlier auxiliary-head
architecture (0.9265 vs.\ 0.7651 previously): roughly 442 non-real clips
are misclassified as real here, versus 1,426 half-truth-as-real
misclassifications reported for the earlier architecture.

\begin{table}[h]
  \centering
  \caption{Per-class precision, recall, and F1 on the unified T2+T3 test
  set (canonical checkpoint, seed 42, $n{=}32{,}800$).}
  \label{tab:perclass}
  \begin{tabular}{lllll}
    \toprule
    \textbf{Class} & \textbf{Precision} & \textbf{Recall} & \textbf{F1} & \textbf{Support} \\
    \midrule
    Real         & 0.9265 & 0.9909 & 0.9576 & 5,600  \\
    Fake         & 0.9829 & 0.9955 & 0.9892 & 11,200 \\
    Half-truth   & 0.9994 & 0.9661 & 0.9825 & 16,000 \\
    \midrule
    Macro avg    & 0.9696 & 0.9842 & 0.9764 & 32,800 \\
    Weighted avg & 0.9813 & 0.9804 & 0.9805 & 32,800 \\
    \bottomrule
  \end{tabular}
\end{table}

\subsection{Feature Contribution Ablation}
\label{sec:ablation}

Table~\ref{tab:ablation} examines feature contributions on MLADDC T2 (5
epochs each, binary track); this concerns feature choice, not the
architectural components of Section~\ref{sec:cafablation}, and is
unaffected by the auxiliary-head decision. Adding MFCC to LFCC improves
accuracy from 96.74\% to 97.96\%. Adding Chroma-STFT under MFAAN's simple
concatenation fusion \emph{reverses} this gain (96.37\%): whether a third
feature axis helps depends on the fusion mechanism's ability to
selectively weight it, motivating cross-attention fusion.

\begin{table}[h]
  \centering
  \caption{Feature contribution ablation on MLADDC T2 (5 epochs each).}
  \label{tab:ablation}
  \begin{tabular}{lll}
    \toprule
    \textbf{Features} & \textbf{Val Acc (\%)} & \textbf{EER (\%)} \\
    \midrule
    MLADDC baseline: LFCC \citep{mladdc2024}       & 68.44 & 40.90 \\
    LFCC only                                     & 96.74 &  2.34 \\
    MFCC + LFCC                                   & 97.96 &  2.27 \\
    MFCC + LFCC + Chroma (MFAAN \citep{mfaan2023}) & 96.37 &  2.21 \\
    \bottomrule
  \end{tabular}
\end{table}

For context, under a fixed, modest fine-tuning budget (5 epochs) matched
to CAFNet's own training regime, two self-supervised baselines fine-tuned
on the same MLADDC T2 binary track, Wav2Vec2 XLS-R (300M parameters,
\citep{xlsr2022}) and the Audio Spectrogram Transformer (87M parameters,
\citep{ast2021}), reach 78.31\% accuracy (4.73\% EER, 0.9901 AUC) and
93.03\% accuracy (7.13\% EER, 0.9810 AUC) respectively, versus CAFNet's
576,027 total parameters, roughly 521$\times$ and 151$\times$ fewer. This
is illustrative context under one fixed, non-exhaustive fine-tuning
budget rather than a ceiling-tuned SSL benchmark, and the two SSL
baselines address only binary detection, not the ternary
classification-plus-localisation task CAFNet's own results above report.

\section{Cross-Dataset Generalisation}
\label{sec:generalisation}

Section~\ref{sec:results} shows performance on data drawn from CAFNet's
training distribution, which does not indicate whether it has learned
anything about partially-manipulated speech in general, as opposed to
MLADDC's specific pipeline, languages, and splice protocol. We apply the
canonical checkpoint zero-shot to two independent benchmarks, evaluating
detection, ternary classification, and localisation separately, since
these need not degrade uniformly, and, as this section shows, need not
respond uniformly to an in-domain architectural change either.

\subsection{Benchmarks}

HAD \citep{halftruth2021} is built on the AISHELL-3 Mandarin corpus and
provides 9,072 test utterances with segment-level labels and explicit
splice timestamps (8,838 half-truth, 234 fully fake). PartialSpoof \citep{zhang2022partialspoof,zhang2021initial,zhang2021mtl}
is built on ASVspoof 2019 LA \citep{wang2020asvspoof2019}; we construct a
balanced 14,000-clip binary evaluation set (7,000 bonafide \texttt{LA\_*},
7,000 spliced \texttt{CON\_*}) from its evaluation partition. Neither
corpus was seen during training.

\subsection{Detection Transfers}
\label{sec:detect}

Table~\ref{tab:zeroshot} reports zero-shot detection on HAD: overall
recall 84.91\% (fake 93.16\%, half-truth 84.69\%), roughly 7.7 points
above the earlier architecture's 77.2\% under the identical protocol.

\begin{table}[h]
  \centering
  \caption{Zero-shot detection transfer to HAD (9,072 clips, canonical
  checkpoint).}
  \label{tab:zeroshot}
  \begin{tabular}{lll}
    \toprule
    \textbf{Class} & \textbf{Recall} & \textbf{n} \\
    \midrule
    Fully fake        & 93.16\% &   234 \\
    Half-truth        & 84.69\% & 8,838 \\
    Overall detection & 84.91\% & 9,072 \\
    \bottomrule
  \end{tabular}
\end{table}

On PartialSpoof, the pattern remains qualitatively different, though it
improves. Table~\ref{tab:partialspoof} shows AUC rising from below chance
(0.4578) to modestly above chance (0.5544), EER falling from 52.53\% to
45.24\%. The TPR/FPR gap narrows from 4.4 to 2.4 points but does not close
and remains inverted: the model still flags genuine audio as ``not real''
slightly \emph{more} often than spliced audio.

\begin{table}[h]
  \centering
  \caption{Zero-shot binary transfer to PartialSpoof (14,000 clips,
  canonical checkpoint). Accuracy is a best-of-99-thresholds upper bound.}
  \label{tab:partialspoof}
  \begin{tabular}{ll}
    \toprule
    \textbf{Metric} & \textbf{Value} \\
    \midrule
    AUC                                    & 0.5544 \\
    EER                                    & 45.24\% \\
    TPR (\texttt{CON\_*} detection, $\theta{=}0.5$)     & 89.89\% \\
    FPR (\texttt{LA\_*} false alarm, $\theta{=}0.5$)    & 92.31\% \\
    \bottomrule
  \end{tabular}
\end{table}

An accuracy figure is deliberately omitted from Table~\ref{tab:partialspoof}:
sweeping the decision threshold and reporting the best value on the same
evaluation set (55.01\%, at $\theta \approx 0.5$) is a threshold-oracle
upper bound rather than a deployable operating point, and is dominated
here by AUC and EER, which are threshold-independent.

Manual inspection of raw softmax outputs on eight PartialSpoof clips
(Table~\ref{tab:qualitative}) confirms the pattern is not a labelling
artefact: genuine \texttt{LA\_*} utterances still receive substantial
``not-real'' probability mass, though the canonical three-way head
sometimes routes that mass to half-truth rather than fake, a distinction
the earlier architecture's more binary-leaning behaviour did not surface.

\begin{table}[h]
  \centering
  \caption{Raw CAFNet softmax output on eight PartialSpoof evaluation
  clips (canonical checkpoint).}
  \label{tab:qualitative}
  \begin{tabular}{llllll}
    \toprule
    \textbf{File} & \textbf{True} & \textbf{P(real)} & \textbf{P(fake)} & \textbf{P(half-truth)} & \textbf{Pred.} \\
    \midrule
    LA\_E\_5996278 & real     & 0.2840 & 0.4097 & 0.3063 & fake \\
    LA\_E\_9305857 & real     & 0.0704 & 0.5345 & 0.3951 & fake \\
    LA\_E\_7484573 & real     & 0.0005 & 0.9955 & 0.0039 & fake \\
    LA\_E\_5504758 & real     & 0.0374 & 0.8821 & 0.0805 & fake \\
    CON\_E\_0042724 & not-real & 0.1963 & 0.4945 & 0.3092 & fake \\
    CON\_E\_0009781 & not-real & 0.0001 & 0.9987 & 0.0013 & fake \\
    CON\_E\_0003269 & not-real & 0.4817 & 0.0995 & 0.4189 & real \\
    CON\_E\_0014452 & not-real & 0.1676 & 0.0449 & 0.7875 & half-truth \\
    \bottomrule
  \end{tabular}
\end{table}

CAFNet was trained exclusively on MLADDC and had never encountered
PartialSpoof's recording conditions, codecs, or splice-construction
protocol during training. Table~\ref{tab:qualitative} shows the model
routes substantial ``not-real'' probability mass onto genuine
\texttt{LA\_*} utterances as readily as onto spliced \texttt{CON\_*}
ones: whatever CAFNet's handcrafted features have learned to treat as
spoofing evidence on MLADDC does not transfer to this corpus.

One plausible confound is duration mismatch rather than a genuine acoustic
failure: MLADDC clips are cropped from 10--20~s sources, so CAFNet's 4~s
window is always fully populated with speech at training time, whereas
PartialSpoof utterances average 3.4~s (69--74\% of clips require
zero-padding, averaging 20--27\% of the window). We tested this directly by
re-extracting features from VAD-trimmed speech-only spans (using
PartialSpoof's own voice-activity annotations, with energy-based trimming
as a fallback) rather than the raw padded clip. AUC fell further under
trimming, from 0.5544 to 0.2925 (EER 45.24\%$\to$65.63\%), ruling out
zero-padding as the dominant cause: the near-chance transfer reflects a
genuine acoustic or channel mismatch, not an input-formatting artefact.

\subsection{Ternary Classification Partially Resolves on HAD}
\label{sec:ternary}

Under the earlier architecture, ternary classification effectively
collapsed to a two-way decision out-of-domain: of 8,838 true half-truth
HAD clips, only 281 (3.2\%) were predicted half-truth, with the model
resolving the rest into fake (72.6\%) or real (24.2\%). Under the
canonical checkpoint, evaluated identically, this no longer holds.
Table~\ref{tab:ternary-had} shows half-truth is now the \emph{modal}
prediction: 4,458 of 8,838 (50.4\%), against 2,881 (32.6\%) fake and
1,499 (16.9\%) real. On the 234 true fake clips, resolution remains
strong: 216 (92.3\%) correctly predicted fake.

\begin{table}[h]
  \centering
  \caption{HAD ternary prediction distribution on 8,838 true half-truth
  clips (canonical checkpoint).}
  \label{tab:ternary-had}
  \begin{tabular}{lll}
    \toprule
    \textbf{Predicted as} & \textbf{n} & \textbf{\%} \\
    \midrule
    half-truth & 4,458 & 50.4\% \\
    fake       & 2,881 & 32.6\% \\
    real       & 1,499 & 16.9\% \\
    \bottomrule
  \end{tabular}
\end{table}

This shift did not follow from any change targeted at HAD or at ternary
classification: the auxiliary head was removed on the strength of
in-domain, single-corpus evidence alone (Section~\ref{sec:cafablation}).
The out-of-domain ternary-resolution rate moved from 3.2\% to 50.4\% as a
side effect of that in-domain-only decision, a change not visible in any
in-domain metric reported in Section~\ref{sec:cafablation}.

\subsection{Localisation Accuracy Out-of-Domain}
\label{sec:oodloc}

\textbf{Protocol.} HAD clips are not natively 4~s, so evaluating the
temporal head requires constructing a 4~s window from a longer utterance
containing the splice. As in Section~\ref{sec:dataset-justification}, the
window's start position is drawn uniformly at random from every offset
for which the window fully contains the splice, but here using HAD's
ground-truth timestamps at test time only, on a checkpoint that never
trained on HAD's own splice-position distribution; we treat this
explicitly as an oracle protocol, since a deployed system would not know
the splice location in advance.

\textbf{Result.} We evaluate on 8,831 of 8,838 HAD half-truth clips (7
excluded for splices longer than 4~s). Table~\ref{tab:hadloc} compares
against the in-domain figure for the same checkpoint. Absolute MAE falls
modestly versus the earlier architecture (0.838~s$\to$0.761~s), but the
\emph{relative} gap to in-domain performance widens, from roughly
11$\times$ to roughly 21$\times$ (0.761/0.037 $\approx$ 20.6), because
in-domain localisation improved by a larger factor than out-of-domain
localisation did. Gated on the model's own ternary classification,
effective end-to-end coverage is now $\sim$50\% of true half-truth clips
(4,458 of 8,838), up from $\sim$3\% (281 of 8,838) under the earlier
architecture, even though the localisation error itself, conditional on
being evaluated, is worse relative to in-domain than before.

\begin{table}[h]
  \centering
  \caption{Boundary localisation: in-domain MLADDC vs.\ zero-shot HAD
  (oracle window), canonical checkpoint.}
  \label{tab:hadloc}
  \begin{tabular}{llll}
    \toprule
    \textbf{Set} & \textbf{MAE (s)} & \textbf{Median (s)} & \textbf{p90 (s)} \\
    \midrule
    MLADDC T3 (in-domain, n=16,000)         & 0.037 & 0.027 & 0.060 \\
    HAD (zero-shot, random-offset, n=8,831) & 0.761 & 0.541 & 1.826 \\
    \bottomrule
  \end{tabular}
\end{table}

\section{Limitations and Conclusion}
\label{sec:conclusion}

\subsection{Limitations}
\label{sec:limitations}

\begin{enumerate}
    \item \textbf{PartialSpoof transfer is unexplained.} Zero-shot transfer
        remains close to chance (Section~\ref{sec:detect}). We ruled out
        input-duration/zero-padding mismatch as the cause (Section~\ref{sec:detect});
        CAFNet's handcrafted-feature front end may instead be picking up
        channel, codec, or recording-condition artefacts specific to MLADDC     
        rather than generalisable spoofing cues; isolating the cause would
        require a channel- and codec-controlled, duration-matched
        re-evaluation, left to future work.
    \item \textbf{Out-of-domain localisation remains an oracle evaluation.}
        A non-oracle protocol (sliding-window search, or a window gated on
        the model's own half-truth classification) is the direct next
        step, and would need to account for effective coverage now being
        $\sim$50\% of true half-truth clips rather than $\sim$3\%.
\end{enumerate}

\subsection{Conclusion}

CAFNet reports, as far as we can establish, the first continuous
splice-boundary localisation result on MLADDC's half-truth track (576,027
parameters, 0.037~s MAE), alongside 97.55\%$\pm$0.69\% ternary accuracy
under 3-seed replication. A component ablation shows cross-attention
fusion is CAFNet's single most load-bearing component and that its
deeply-supervised auxiliary classification head is not: removing it
improves every in-domain metric, including per-class precision on the
weakest (real) class, with lower run-to-run variance, and we adopt the
auxiliary-head-free architecture as canonical.

The cross-dataset study built on that canonical checkpoint shows two
things. First, the same asymmetric, capability-specific transfer pattern
found previously persists: coarse detection transfers well to HAD and
poorly to PartialSpoof, ternary classification and detection are
separable capabilities that do not degrade together, and localisation
remains the least robust output in relative terms. Second, an
architectural change validated purely on in-domain metrics shifted that
cross-corpus transfer profile in different directions for different
outputs: detection and ternary classification on HAD improved
substantially, in the case of ternary classification from a near-total
collapse to a roughly even split, while localisation's relative
out-of-domain degradation widened even as its absolute error improved.

Future work follows directly from Section~\ref{sec:limitations}: a
controlled cross-corpus diagnostic study to establish why PartialSpoof
transfer remains close to chance, and a non-oracle localisation protocol
for out-of-domain evaluation.
\section*{Statements and Declarations}

\medskip\noindent
\textbf{Declaration of competing interest:}
The authors declare that they have no known competing financial interests
or personal relationships that could have appeared to influence the work
reported in this paper.

\medskip\noindent
\textbf{Funding:}
The authors received no specific funding for this work.

\medskip\noindent
\textbf{Data Availability:}
MLADDC T2 \& T3: \url{https://kaggle.com/datasets/artharking/mladdc-t2} \&
\url{https://kaggle.com/datasets/artharking/mladdc-t3}. HAD and PartialSpoof
are publicly available from their respective original publications
\citep{halftruth2021,zhang2022partialspoof}.

\medskip\noindent
\textbf{Code Availability:}
Code and trained models are publicly available at
\url{https://github.com/ssutharya/Audio_Deepfake_Detection}.

\medskip\noindent
\textbf{CRediT authorship contribution statement:}
\textbf{S. Sutharya:} Conceptualization, Methodology, Software, Validation,
Formal analysis, Investigation, Data curation, Writing - original draft,
Visualization. \textbf{Remya K Sasi:} Supervision, Writing - review \& editing.

\section*{Declaration of Generative AI and AI-Assisted Technologies in the
Writing Process}
During the preparation of this work the authors used Claude (Anthropic) to
assist with condensing, structuring, and refining manuscript text, and to
improve academic readability and tone. After using these tools, the authors
reviewed and edited the content as necessary and take full responsibility
for the content of the published article.

\bibliographystyle{elsarticle-num}

\end{document}